\documentclass[twocolumn,prd,aps,superscriptaddress]{revtex4}
\usepackage{aas_macros}
\usepackage{dcolumn}
\usepackage{bm}
\usepackage{amsmath,amssymb}
\usepackage{color}
\usepackage{graphicx}
\usepackage{ulem}

\def\der{{\rm d}}

\newcommand{\be}{\begin{equation}}
\newcommand{\ee}{\end{equation}}
\newcommand{\ba}{\begin{eqnarray}}
\newcommand{\ea}{\end{eqnarray}}

\begin{document}
\title{Testing interacting dark matter and dark energy model with cosmological data}

\author{Gong Cheng}
\affiliation{Key Laboratory of Computational Astrophysics, 
National Astronomical Observatories, Chinese Academy of Sciences,
20A Datun Road, Beijing 100101, China}
\affiliation{School of Astronomy and Space Science, University of Chinese Academy of Sciences, Beijing 100049, China}

\author{Yin-Zhe Ma}
\affiliation{School of Chemistry and Physics, University of KwaZulu-Natal, Westville Campus, Private Bag X54001, Durban, South Africa}
\affiliation{NAOC-UKZN Computational Astrophysics Center (NUCAC), University of Kwazulu-Natal, Durban, 4000, South Africa}

\author{Fengquan Wu}
\affiliation{Key Laboratory of Computational Astrophysics, 
National Astronomical Observatories, Chinese Academy of Sciences,
20A Datun Road, Beijing 100101, China}

\author{Jiajun Zhang}
\affiliation{Center for Theoretical Physics of the Universe, Institute for Basic Science (IBS), Daejeon, 34126, Korea}

\author{Xuelei Chen}
\affiliation{Key Laboratory of Computational Astrophysics, 
National Astronomical Observatories, Chinese Academy of Sciences,
20A Datun Road, Beijing 100101, China}
\affiliation{School of Astronomy and Space Science, University of Chinese Academy of Sciences, Beijing 100049, China}
\affiliation{Center for High Energy Physics, Peking University, Beijing 100871, China}

\date{\today}

\begin{abstract}
We investigate the model of dark matter-dark energy (DM-DE) interaction with coupling strength proportional to the multiplication of dark sector densities with different power indices $Q = \gamma \rho_{\rm c}^{\alpha} \rho_{\rm d}^{\beta}$. We first investigate the modification of the cosmic expansion history, and then further develop the formalism to take into account the cosmological perturbations and dark matter temperature evolution. We then use the latest observational
cosmology data, including cosmic microwave background (CMB) data, baryon acoustic oscillations (BAO) data, 
redshift-space distortion (RSD) data and Type Ia supernovae (SNe) data to constrain the model parameters. We find in the phantom region, a positive $\alpha$ is preferred by the data above $2\, \sigma$ statistic significance. If we choose the power indices to be integers 
or half-integers for {\it plausible} physics of particle interaction, the allowed values within $1\, \sigma$ confidence regions are $\alpha = 0.5$ 
and $\beta = 0, 0.5, 1$. The inclusion of BAO and RSD data from large-scale structure and SNe data improves the constraints significantly. Our model predicts lower values of $f(z) \sigma_8(z)$ at $z<1$ comparing to $\Lambda$CDM model, which alleviates the tension of $\Lambda$CDM with various RSD data from optical galaxy surveys. Overall, the DM-DE interaction model is consistent with the current observational data, especially providing a better fit to the RSD data.
\end{abstract}


\maketitle

\section{INTRODUCTION}
Recently, the {\it Planck} measurement of the cosmic microwave background radiation (CMB) produces the best-fitting cosmological parameters, which are in good agreement with the low 
redshift observations (e.g., Baryon Acoustic Oscillation (BAO) from galaxy survey, Type-Ia supernovae data (SNe) and galaxy lensing measurements)~\cite{2018arXiv180706209P}. However, 
there is a roughly $2.5\, \sigma$ confidence level (C.L.) tension between the base {\it Planck} $\Lambda$CDM cosmology prediction and 
the DES combined-probe results (the later prefers a lower late-time clustering amplitude $\sigma_8$ or matter 
density $\Omega_m$). Besides, the {\it Planck} base $\Lambda$CDM cosmology requires 
$H_0 = (67.4 \pm 0.5)\,{\rm km}\,{\rm s}^{-1}\,{\rm Mpc}^{-1}$, which leads to a $4.0\, \sigma$--$5.8\, \sigma$ tension 
with the local Hubble constant measurements depending on the approaches used \cite{Verde2019}. 
Many new observations of Hubble constant and new physics trying to explain 
the discrepancy are discussed in Ref.~\cite{Verde2019}.

Theoretically, new ideas beyond $\Lambda$CDM model are developed, aiming to solve the well-known cosmological 
constant problem and coincidence problem. Moreover, it has been pointed out that the effective field theory that 
compatible with string theory should satisfy the swampland criteria, while the cosmological constant scenario 
does not \cite{Obied2018}. 

One possible way to alleviate these problems is to consider the interaction between dark sectors \cite{Santos2017}. 
It is reported that the $H_0$ and $\sigma_8$ tension could be solved or reduced simultaneously by considering an 
interaction in the dark sector \cite{Kumar2019, Valentino2019, 2018JCAP...09..019Y, 2019arXiv191009853D}. 
Also, it could alleviate the coincidence problem by allowing 
a constant ratio of dark sector densities \cite{Costa2017}. In Ref.~\cite{Bruck2019}, the authors explore the possibility that the 
interaction functions possess a minimum and argue that this model could alleviate the tension between the swampland 
conjectures and the quintessential potential \cite{Bruck2019}. Besides, in the framework of field theory, it is natural and inevitable to 
consider such interactions between dark sectors, and the investigation of the interaction could help us to understand the 
nature of them \cite{Costa2014}.

Various models are proposed and tested in the literature (for reviews, see \cite{WangBin2016, Wang2017, Bolotin2015}).
A much studied case is $Q = H (\xi_1 \rho_{\rm c} + \xi_2 \rho_{\rm d})$. Other models are also considered, such as 
$Q = \Gamma (\dot{\rho}_{\rm c} + \dot{\rho}_{\rm d})$~\cite{Shahalam2015}, $Q = H \Gamma \rho_{\rm c} \rho_{\rm d} / (\rho_{\rm c} + \rho_{\rm d})$ 
\cite{Campo2015}, where $\rho_{\rm c}$ and $\rho_{\rm d}$ are the energy densities of dark matter and dark energy respectively and 
$\Gamma, \xi_1, \xi_2$ describe the interaction strength. Besides, the holographic principle is applied in some work 
\cite{Ma2010, Wang2017} and the coupled quintessence model is discussed in Ref.~\cite{Mifsud2017}.

However, a physically more plausible form of interaction form is for the interaction term to be proportional to the product 
of the densities of interacting components or some powers of these \cite{Mangano2003, Ma2010}. For example, 
in the familiar case of chemical reactions this is the case. 
Thus a natural way to construct the phenomenological model is to suppose the interaction 
term is proportional to some powers of the densities,
\ba
\label{eq:back1}
\dot{\rho}_{\rm c} + 3 a H \rho_{\rm c} &=& a \gamma \rho_{\rm c}^{\alpha} \rho_{\rm d}^{\beta},\\
\label{eq:back2}
\dot{\rho}_{\rm d} + 3 a (1 + w) H \rho_{\rm d} &=& - a \gamma \rho_{\rm c}^{\alpha} \rho_{\rm d}^{\beta},
\ea
where the dot denotes the time derivative with respect to the conformal time and a is the scale factor. 
In this work,  we consider a minimal model in which the equation of state of dark energy $w$ is a constant. 
The power index $\alpha, \beta$ are assumed to be non-negative numbers. 
If we set the indices to be simple integers, the model could reduce to the interactions in the usual sense, such as 
``decay" (if one is 0 and the other is 1), ``annihilation" (if one is 0, and the other is 2). And $\gamma > 0$ indicates dark 
energy is converted to dark matter, while $\gamma < 0$ indicates the opposite process.

Previous work on the interacting dark energy models mainly focuses on the impact on the background evolutions and 
expansion history. However, from the theoretical point of view, the inhomogeneity of dark matter will naturally lead to 
dark energy perturbations as the energy transfers between the two components. And the change of density and velocity 
perturbations will influence the structure formation and some observational effects (e.g., redshift space distortion). 
So in this paper, we also investigate the impact on the growth of cosmological perturbations. 

Notably, as the perturbed interaction 
four-vector $\delta Q^{\nu}$ cannot be determined uniquely by the background interaction form $Q$, the 
formalisms developed by different groups differ from each others slightly. In Refs.~\cite{Gavela2009, Lopez2010, Gavela2010}, 
$\delta Q^{\nu} = \delta Q u^{\nu}_{\rm c}/a$ or $\delta Q^{\nu} = \delta Q u^{\nu}_{\rm d}/a$ is assumed to avoid momentum transfer in the rest 
frame of dark sector, where $u^{\nu}_{\rm d}$ and $u^{\nu}_{\rm d}$ are the four velocities. In Refs.~\cite{He2011, Costa2014}, on the 
other hand, the authors assume the energy transfer is stationary and non-gravitational interaction between dark sectors does 
not exist and so the non-vanishing component is $\delta Q^{0} = \delta Q /a$. In this paper, we studied the problem with a specific Lagrangian model, for which 
$\delta Q^{i}_{(\lambda)} = Q^0_{(\lambda)} v_{(\lambda)} (\lambda = {\rm c}, {\rm d})$. However, the different choices of $\delta Q^{i}$ have 
negligible influence on the observables, only $\delta Q^{0}$ matters in the perturbation evolutions.

\section{Formalisms}
In this section, we present the formalisms of background evolutions, linear perturbations and thermodynamics in this model.
The background evolutions of dark matter and dark energy are described by Eqs.~(\ref{eq:back1}) and (\ref{eq:back2}). Due 
to the energy transfer, the energy-momentum tensor of each component is no longer conserved,
\be
\nabla_{\mu}T^{\mu\nu}_{(\lambda)} = Q^{\nu}_{(\lambda)}.
\ee
We can specify the background coupling vector $Q^{\nu}_{(\lambda)}$ in the comoving frame as \cite{He2011}
\ba
Q_{(\lambda)}^{\nu} &=& \left[\frac{Q_{(\lambda)}}{a},0,0,0 \right]^T, \\
Q_{(\lambda)}^{\nu} Q_{\nu (\lambda)} &=& g_{00} \left(Q_{(\lambda)}^0\right)^2 = - Q_{(\lambda)}^2, \label{eq:module}\\
Q &=& Q_{\rm c} = - Q_{\rm d} = \gamma \rho_{\rm c}^{\alpha} \rho_{\rm d}^{\beta}, \label{eq:interact}
\ea where $Q_{(\lambda)}$ is the module of $Q_{(\lambda)}^{\nu}$, and $Q$ is the energy transfer rate with respect to the cosmic time. 
For convenience, we use the critical density today to define the dimensionless 
interaction parameter as
\be
\lambda = \gamma \rho_{\rm cr}^{\alpha +\beta -1} H_0^{-1}, ~~\rho_{\rm cr} = 3 H_0^2 M_{\rm pl}^2.
\ee

For the linear perturbations in the presence of the interaction, we first review the formalisms developed in Refs. 
\cite{Ma1995, He2011, Costa2014}, and then apply them in this model. 
To obtain the zero component of the perturbed energy-momentum transfer, one can perturb Eq.~(\ref{eq:module}) and find
\be
\delta Q_{(\lambda)}^{0} = -\frac{\psi}{a} Q_{(\lambda)} + \frac{1}{a} \delta Q_{(\lambda)}, \label{eq:q0}
\ee
where $\psi$ is the scalar metric perturbation. $\delta Q_{(\lambda)}^{0}$ is uniquely determined by the background coupling vector 
$Q^{\nu}_{(\lambda)}$ and it is proved that it is covariant \cite{He2011}. However, the spatial component of the perturbed 
energy-momentum transfer $\delta Q^i_{(\lambda)}$ has to be specified based on the physics. The potential of 
$\delta Q^i_{(\lambda)}$ could be decomposed as
\be
\delta Q_{{\rm p}(\lambda)}=\left.Q_{{\rm p}(\lambda)}^{I}\right|_{\rm t}+Q_{(\lambda)}^{0} v_{\rm t}.
\ee
$\left.Q_{{\rm p}(\lambda)}^{I}\right|_{\rm t}$ denotes the external non-gravitational force density between the two components and 
$v_t$ is the energy transfer velocity. As mentioned in the Introduction, $v_{\rm t}$ is set to be 0 or follows the velocities of dark sectors in the
literature. 

In the following analysis, we illustrate how to determine $\delta Q^{i}$ in a concrete example of our phenomenological model.
We assume dark matter and dark energy are described by two scalar fields $\psi$ and $\phi$, respectively. The Lagrangian reads
\ba
\nonumber L =&& \sqrt{-g} \bigg[ \frac{R}{16 \pi G} - \frac{1}{2}\nabla^{\mu} \psi \nabla_{\mu} \psi - V(\psi) - \frac{1}{2}\nabla^{\mu} \phi \nabla_{\mu} \phi 
\\  &&- U(\phi) - I(\psi,\phi) \bigg].
\ea
If we assume the interaction term $I(\psi, \phi)$ and the scalar potential as follows, Eq.~(\ref{eq:interact}) is recovered.
\ba
I(\psi, \phi) &=& \frac{\gamma C^{2\alpha + 2\beta -1}}{4^{\alpha+\beta} \alpha^{2 \alpha} (1+w)^\beta \beta^{2 \beta}} \psi^{2\alpha} \phi^{2\beta},\\
V(\psi) &=& \frac{C^2}{8\alpha^2} \psi^2,\\
U(\phi) &=& \frac{C^2}{8\beta^2} \frac{1-w}{1+w}\phi^2,
\ea
where $C = Q / I(\psi,\phi)$ is a constant in unit of $s^{-1}$.

The field equation and the equations of motion for the scalar fields are
\ba
G_{\mu \nu} = 8\pi G\big[T_{\mu \nu}^{({\rm c})} &+& T_{\mu \nu}^{({\rm d})} -g_{\mu \nu} I(\psi,\phi)\big],\\
\Box \psi &-&V_{,\psi} - I_{,\psi} = 0,\\
\Box \phi &-&V_{,\phi} - I_{,\phi} = 0,
\ea
where
\ba
\nonumber T_{\mu \nu}^{({\rm c})} &=&\nabla_{\mu} \psi \nabla_\nu \psi - g_{\mu \nu}\left[\frac{1}{2}g^{\rho \sigma} \nabla_\rho \psi \nabla_\sigma \psi +V(\psi)\right],\\
\\
T_{\mu \nu}^{({\rm d})} &=&\nabla_{\mu} \phi \nabla_\nu \phi - g_{\mu \nu}\left[\frac{1}{2}g^{\rho \sigma} \nabla_\rho \phi \nabla_\sigma \phi +U(\phi)\right].
\ea
So we obtain
\ba
\nabla_\mu T^{\mu \nu}_{({\rm c})} &=& I_{,\psi} \partial^\nu \psi = Q^\nu_{({\rm c})},\\
\nabla_\mu T^{\mu \nu}_{({\rm d})} &=& I_{,\phi} \partial^\nu \phi = Q^\nu_{({\rm d})}.
\ea
$\delta Q_{(\lambda)}^{0}$ is consistent with Eq.~(\ref{eq:q0}) and $\delta Q^{i}_{(\lambda)} = Q^0_{(\lambda)} v_{(\lambda)}$.
However, we have checked that the inclusion of $\delta Q^{i}$ has negligible impact on the cosmological observables. 

In the synchronous gauge, the density and velocity perturbation equations for dark matter and dark energy in our model read 
\ba
\dot{\delta}_{\rm c} &=& -(\theta_{\rm c} + \frac{\dot{h}}{2}) + a \gamma \rho_{\rm c}^{\alpha-1} \rho_{\rm d}^{\beta} [(\alpha-1) \delta_{\rm c} + \beta \delta_{\rm d}],\\
\dot{\theta}_{\rm c} &=& -a H \theta_{\rm c} - a \gamma \rho_{\rm c}^{\alpha - 1} \rho_{\rm d}^{\beta} \theta_{\rm c} + k^2 c_{\rm c}^2 \delta_{\rm c},\\
\nonumber \dot{\delta}_{\rm d} &=& -(1 + \omega) (\theta_{\rm d} + \frac{\dot{h}}{2}) + 3 a H (\omega - c_{\rm e}^2)\delta_{\rm d}\\
\nonumber && - a \gamma \rho_{\rm c}^{\alpha} \rho_{\rm d}^{\beta - 1}[(\beta - 1) \delta_{\rm d} + \alpha \delta_{\rm c}] -3 a H\\
&&(c_{\rm e}^2 - c_a^2)[3 a H (1 + \omega) + a \gamma \rho_{\rm c}^{\alpha}\rho_{\rm d}^{\beta -1}]\frac{\theta_{\rm d}}{k^2},\\
\nonumber \dot{\theta}_{\rm d} &=& -a H \theta_{\rm d} (1-3c_{\rm e}^2) - \frac{\dot{w}}{1+w}\theta_{\rm d} + \frac{1}{1+w}\\
&&(1+c_{\rm e}^2)a\gamma \rho_{\rm c}^{\alpha}\rho_{\rm d}^{\beta-1}\theta_{\rm d}+\frac{k^2c_{\rm e}^2\delta_{\rm d}}{1+w},
\ea
where we have used the form of the perturbed pressure of dark energy in a general frame \cite{Xia2008, Costa2014},
\be
\delta P_{\rm d}=c_{\rm e}^{2} \delta_{\rm d} \rho_{\rm d}+\left(c_{\rm e}^{2}-c_{a}^{2}\right)\left[\frac{3 aH(1+\omega) \theta_{\rm d} \rho_{\rm d}}{k^2}
-a^{2} Q_{\rm d}^{0} \frac{\theta_{\rm d}}{k^2}\right].
\ee
In practice, we set the effective sound speed of dark energy in the rest frame $c_{\rm e} = 1$ and adiabatic sound speed of dark 
energy $c_a^2 = w$.

We set the adiabatic initial conditions for the perturbations following \cite{Ballesteros2010}. In order to obtain analytical 
solutions of initial conditions, we neglect all the interaction terms when solving the continuity and Euler equations. 
So our initial conditions are identical with Ref.~\cite{Ballesteros2010}.
\ba
\delta_{\rm c} &=& \frac{3}{4} \delta_{\gamma} = -\frac{1}{2} C (k \tau)^2,\\
\theta_{\rm c} &=& 0,\\
\theta_{\gamma} &=& -\frac{1}{18}C(k^4\tau^3),\\
\nonumber \delta_{\rm d} &=& -\frac{C}{2}(1+w)\frac{4-3c_{\rm e}^2}{4-6w+3c_{\rm e}^2}(k\tau)^2 \\
&=& \frac{1+w}{7-6w}\delta_{\rm c},\\
\theta_{\rm d} &=& -\frac{C}{2}\frac{c_{\rm e}^2}{4-6w+3c_{\rm e}^2}(k\tau)^3k = \frac{9}{7-6w}\theta_{\gamma},
\ea
where $\gamma$ corresponds to photons and $C$ is a constant.

Besides, the term $k^2 c_{\rm c}^2 \delta_{\rm c}$ appearing in the RHS of $\dot{\theta}_{\rm c}$ is always ignored by previous work. 
Nevertheless, compared with $\Lambda$CDM model, $\delta_{\rm c}$ and $c_{\rm c}$ are affected by the interaction. Hence 
this term might be enhanced and should be taken into account. It's necessary to investigate the contribution of this term 
to $\dot{\theta}_{\rm c}$. The sound speed of dark matter $c_{\rm c}$ is defined as~\cite{Ma1995}
\be
c_{\rm c}^2 = \frac{k_{\rm B} T_{\rm c}}{m_{\rm c}} \left(1 - \frac{{\rm d} \ln T_{\rm c}}{3 {\rm d} \ln a}\right),
\ee
where $m_{\rm c}$ is the mass of dark matter particle.
We follow the methods in Refs.~\cite{Maartens1996, Cardenas2019} to calculate the temperature evolution of dark matter 
in the interacting model using the second law of thermodynamics and obtain
\ba
\nonumber \dot{T_{\rm c}} &=& \left(\frac{\partial T_{\rm c}}{\partial n_{\rm c}}\right)_{\rho_{\rm c}} \dot{n_{\rm c}} + \left(\frac{\partial 
T_{\rm c}}{\partial \rho_{\rm c}}\right)_{n_{\rm c}} \dot{\rho_{\rm c}}\\
\nonumber &=& \left(\frac{\partial T_{\rm c}}{\partial n_{\rm c}}\right)_{\rho_{\rm c}}\left(-3aHn_{\rm c} + \frac{aQ}{m_{\rm c}}\right)\\
\nonumber&&+ \left(\frac{\partial T_{\rm c}}{\partial \rho_{\rm c}}\right)_{n_{\rm c}}\left(-3aH \rho_{\rm c} + aQ\right) \\
\label{eq:T}
&=& -2 a H T_{\rm c} \left(1 - \frac{\gamma \rho_{\rm c}^{\alpha -1} \rho_{\rm d}^{\beta}}{3 H}\right),
\ea
where we have used $T\left(\frac{\partial p}{\partial \rho}\right)_n = (\rho + p)\left(\frac{\partial T}{\partial \rho}\right)_n 
+ n\left(\frac{\partial T}{\partial n}\right)_{\rho}$, $\left(\frac{\partial p_{\rm c}}{\partial \rho_{\rm c}}\right)_{n_{\rm c}} = 2/3$ and $n$ is the number density. 
As the wrong relation $\left(\frac{\partial p_{\rm c}}{\partial \rho_{\rm c}}\right)_{n_{\rm c}} = w_{\rm eff} = w_c - \frac{Q}{3H\rho_{\rm c}}$ is used 
in Ref.~\cite{Cardenas2019}, Eq.~(34) in Ref.~\cite{Cardenas2019} is related to Eq.~(\ref{eq:T}) in this paper by a factor $2/3$.

To solve the above equation numerically, we need to set the initial temperature in the early universe. If we assume 
$T_{\rm c} = 0$ initially, we obtain a trivial solution, as the macroscopic interaction of the dark matter 
does not generate the microscopic motion and temperature. Here we set the initial temperature as evolved from earlier time, 
assuming that dark matter annihilates to baryons through weak-scale interactions at high energies. 
The freeze-out redshift can be obtained by \cite{Xu2018}
\be
T_{\rm c}(z)=T_{\rm b}(z) \quad \text { at } \quad H(z)= \left\langle\sigma_{w} v\right\rangle \rho_{\rm c} (z)/ m_{\rm c},
\ee
where we take the weak-scale cross section $\left\langle\sigma_{w} v\right\rangle \sim 10^{-26}\,{\rm cm}^3\,{\rm s}^{-1}$ and $T_{\rm b}$ 
is the baryons temperature. After freeze-out, $T_{\rm c}(z)$ evolves adiabatically until being influenced by the DM-DE interaction at low redshifts.

\section{Methods}
We modify the public Boltzmann code {\tt CLASS} \cite{Lesgourgues2011, Blas2011} to implement our model and to compute 
the theoretical values of the observables. Given the present-day Hubble parameter $H_0$ and the fraction of dark matter 
$\Omega_{\rm c}$, we can use the shooting method to obtain the initial conditions of $\rho_{\rm c}$ and $\rho_{\rm d}$. 
Note that we only assume dark energy interacts with dark matter, so the evolution of baryons is unchanged. 
Then we use the code {\tt Monte Python} \cite{Audren2013, Brinckmann2018} which adopts the Markov chain Monte Carlo (MCMC) method 
to constrain the parameters in this model by fitting the cosmological data.

The data we have used includes CMB data, BAO data, redshift space distortion (RSD) data and SNe data. The CMB data consists of {\it Planck} 2015 
temperature, polarization power spectrum (TT, TE, EE, low-$\ell$) and lensing measurements \cite{2016A&A...594A...1P, 2016A&A...594A..13P}. 
We also combine recent and reliable BAO and RSD measurements from various surveys, summarized in Tables~ 
\ref{table:bao} and \ref{table:rsd}. As pointed out in Ref.~\cite{2018arXiv180706209P}, Quasar Ly$\alpha$ measurements 
are based on some assumptions and WiggleZ survey volume overlaps with BOSS-CMASS partly. So we do not 
include the BAO data from Quasar Ly$\alpha$ and WiggleZ in our analysis.

\begin{table}
\begin{center}
\caption{\label{table:bao}BAO measurements from various surveys adopted in this work.}
\begin{tabular}{cccc}
 \hline  \hline
Redshift & Measurement & Value & Surveys\\ \hline
0.106 & $r_{\rm s}/D_{\rm V}$ & $0.327 \pm 0.015$ & 6dFGS \cite{6dF-bao}\\
0.15 & $D_{\rm V}/r_{\rm s}$ & $4.47 \pm 0.16$ & SDSS DR7-MGS \cite{MGS-bao}\\
0.35 & $D_{\rm V}/r_{\rm s}$ & $9.11 \pm 0.33$ & SDSS DR7-LRG \cite{LRG-bao}\\
0.38 & $D_{\rm M} (r_{\rm s,fid}/r_{\rm s})$ & $1518.4 \pm 22.4$ & SDSS DR12-BOSS \cite{dr12-combined}\\
0.38 & $H(z) (r_{\rm s}/r_{\rm s,fid})$ & $81.51 \pm 1.91$ & SDSS DR12-BOSS\\
0.51 & $D_{\rm M} (r_{\rm s,fid}/r_{\rm s})$ & $1977.4 \pm 26.5$ & SDSS DR12-BOSS\\
0.51 & $H(z) (r_{\rm s}/r_{\rm s,fid})$ & $90.45 \pm 1.94$ & SDSS DR12-BOSS\\
0.61 & $D_{\rm M} (r_{\rm s,fid}/r_{\rm s})$ & $2283.2 \pm 31.9$ & SDSS DR12-BOSS\\
0.61 & $H(z) (r_{\rm s}/r_{\rm s,fid})$ & $97.26 \pm 2.09$ & SDSS DR12-BOSS\\
1.52 & $D_{\rm V}/r_{\rm s}$ & $26.005 \pm 0.995$ & SDSS DR14\cite{dr14-bao}\\
\hline
\end{tabular}
\end{center}
\end{table}

\begin{table}
\begin{center}
\caption{\label{table:rsd}RSD measurements from various surveys adopted in this work.}
\begin{tabular}{ccc}
 \hline  \hline
Redshift & $f(z) \sigma_8(z)$ & Surveys\\ \hline
0.02 & $0.428 \pm 0.0465$ & Velocities from SNe \cite{sn-bao}\\
0.067 & $0.423 \pm 0.055$ & 6dFGS \cite{6df-rsd}\\
0.15 & $0.49 \pm 0.15$ & SDSS DR7-MGS \cite{MGS-rsd}\\
0.3 & $0.49 \pm 0.09$ & SDSS DR7-LRG \cite{lrg-rsd}\\
0.18 & $0.36 \pm 0.09$ & GAMA \cite{gama}\\
0.38 & $0.44 \pm 0.06$ & GAMA\\
0.38 & $0.4975 \pm 0.0451$ & SDSS DR12-BOSS\\
0.51 & $0.4575 \pm 0.0377$ & SDSS DR12-BOSS\\
0.61 & $0.4361 \pm 0.0344$ & SDSS DR12-BOSS\\
0.44 & $0.435 \pm 0.055$ & WiggleZ \cite{wigglez-rsd}\\
0.60 & $0.451 \pm 0.042$ & WiggleZ\\
0.73 & $0.478 \pm 0.038$ & WiggleZ\\
0.6 & $0.55 \pm 0.12$ & VIPERS \cite{vipers}\\
0.86 & $0.40 \pm 0.11$ & VIPERS\\
1.4 & $0.482 \pm 0.116$ & FastSound \cite{fastsound}\\
1.52 & $0.426 \pm 0.077$ & SDSS DR14 \cite{dr14-rsd}\\
\hline
\end{tabular}
\end{center}
\end{table}

Table~\ref{table:bao} shows the measurements of distance ratio $D_{\rm V}/r_{\rm s}$ at the effective redshift.
$D_{\rm V}$ is a combination of Hubble parameter $H(z)$ and comoving angular diameter distance $D_{\rm M}(z)$ 
\cite{2018arXiv180706209P}, 
\be
D_{\rm V} (z) = \left[D_{\rm M}^2(z) \frac{c z}{H (z)} \right]^{1/3}.
\ee
$r_{\rm s}$ is the comoving sound horizon at the end of the baryon drag epoch.

The peculiar velocity of galaxies could lead to distortion of clustering of galaxies in the redshift space. 
So measuring RSD effect can help us to probe the growth function. Table~\ref{table:rsd} shows the constraints on $f(z) \sigma_8(z)$
from various surveys. The scale-independent growth function is defined as
\be
f(z) = \frac{{\rm d} \ln D}{{\rm d} \ln a}, \,\,\,\, D(a) = \frac{\delta (a)}{\delta (a_0)}.
\ee

We use the SNe data from the ``Pantheon Sample", which consists of 1048 SNe with the redshift spanning $0.01 < z < 2.3$. 
This sample is a combination of SNe Ia from Pan-STARRS1 Medium Deep Survey, SNLS, and several low-$z$ and Hubble 
Space Telescope samples \cite{pantheon}.

Due to the divergency of perturbations when $w$ approaches $-1$, we constrain the models with $w<-1$ and $w>-1$ separately.  
We choose a broad range for the dark matter mass $m_{\rm c}$, and we do not impose
boundary on $\lambda$, both positive and negative values are allowed. 
We impose the following upper bound on the power indices $\alpha$ and $\beta$. Physically, if the interaction is due to 
coupling between two fields,  one might expect a small value of $\alpha$ and $\beta$ from their interaction, while a single term of large power 
is not very plausible. Also,  it is necessary to set some prior range for the computation in practice. After some trials,  
 The prior ranges are set as
\ba
0< \alpha &<& 40,\\
0< \beta &<& 40,\\
-2 < w &<& -1~~~ \text{or}~~~ -1< w < -0.5,\\
1 ~\rm eV &<& m_{\rm c} < 10 ~\rm TeV.
\ea
We shall discuss these priors further below.

\section{Results and Discussions}

\begin{figure*}[htb]
\includegraphics[width=0.5\textwidth]{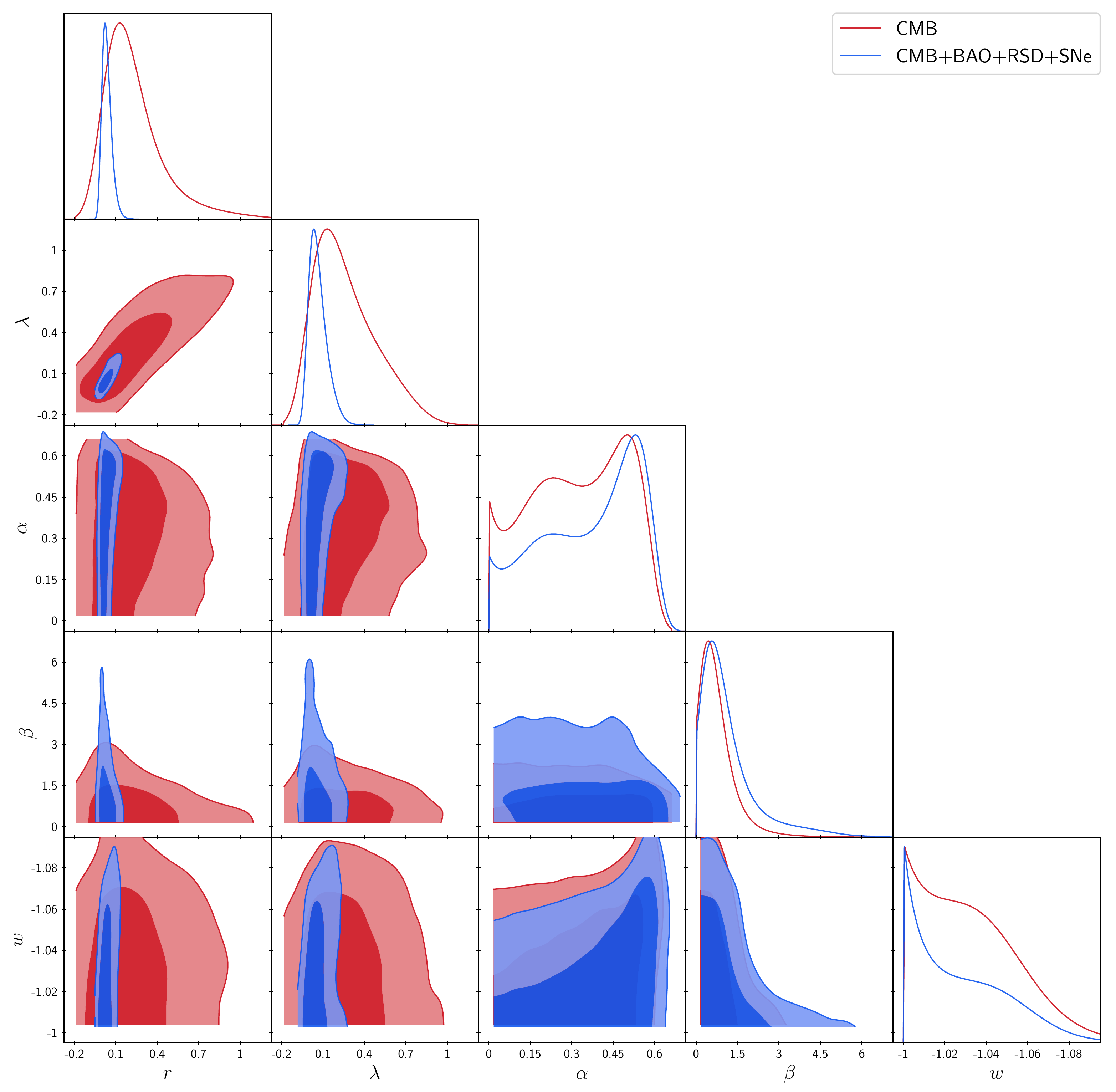}\includegraphics[width=0.5\textwidth]{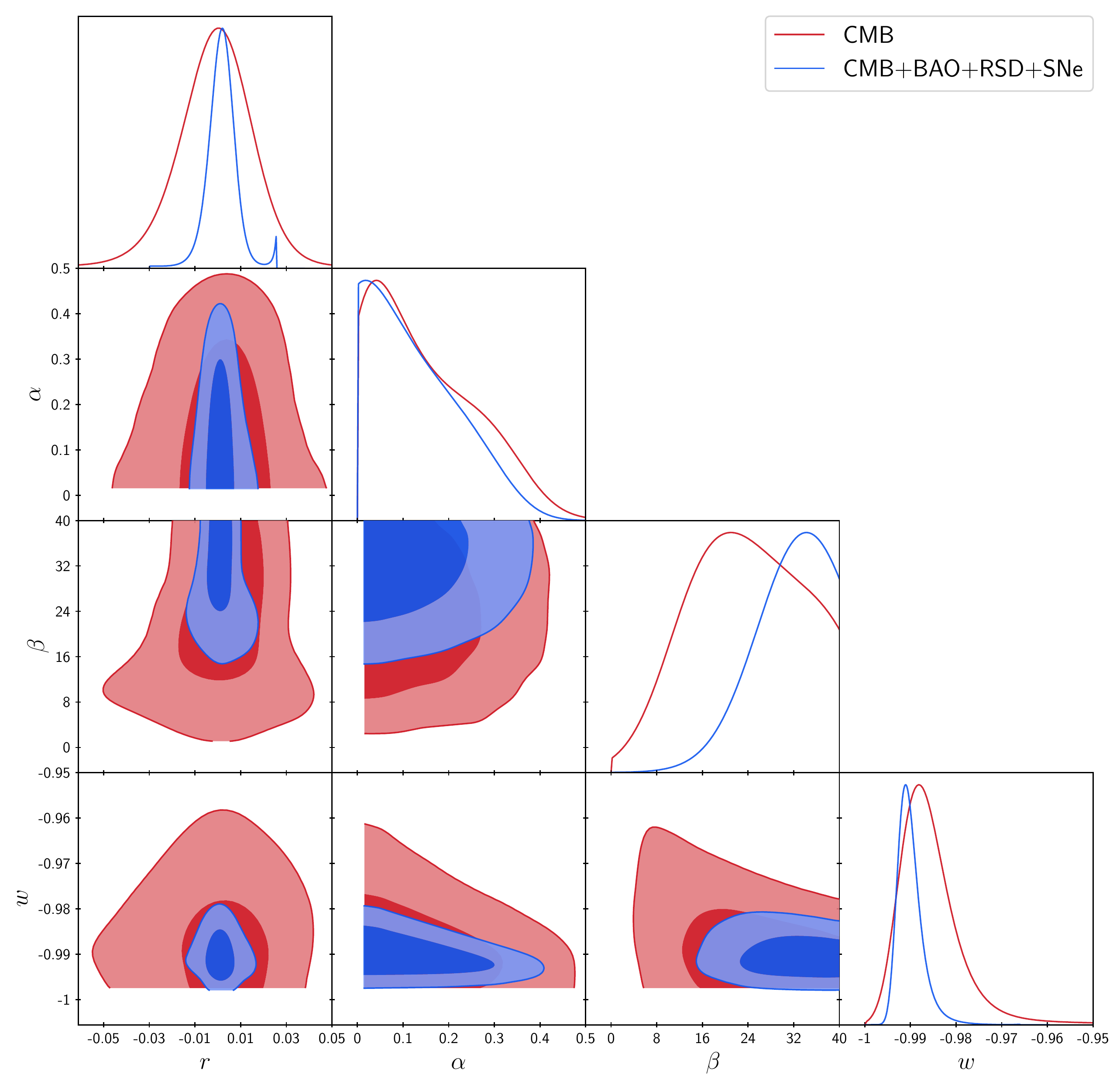}
\caption{One-dimensional marginalized posterior distribution and $68\%$, $95\%$ confidence regions of the four free parameters in 
the model. The red lines correspond to the constraining results from CMB data only, and the blue lines correspond to joint 
constraints from CMB+BAO+RSD+SNe dataset. {\it Left} panel is for the model $w<-1$ and the {\it right} panel is for $w>-1$.
To illustrate the strength of interaction, we also plot the constraints for $r$, defined as $r = a \gamma \rho_{\rm c}^{\alpha} \rho_{\rm d}^{\beta} / 
(a H \rho_{\rm d}) = \lambda \Omega_{\rm c}^\alpha \Omega_{\rm d}^{\beta-1}$, which is the ratio between interaction term and 
expansion term in Eq.~(\ref{eq:back2}) at present day. For the $w>-1$ case, we do not plot $\lambda$ contours as it is not well constrained.
}
\label{fig:constraints}
\end{figure*}

\begin{table*}
\begin{center}
\scriptsize
\caption{\label{table:result1}Constraints on the parameters in the model $w<-1$.}
\begin{tabular}{cccccc}
\hline  \hline
& {\it Planck} & {\it Planck} & ~{\it Planck}+BAO+RSD+SNe~ & ~{\it Planck}+BAO+RSD+SNe~ & ~{\it Planck}+BAO+RSD+SNe~ \\
Parameter & $68\%$ C.L. & $95\%$ C.L. & $68\%$ C.L. & $95\%$ C.L. & Best fit \\ \hline\\ \vspace{0.5em}
$\lambda$ & $0.27_{-0.29}^{+0.15}$ & $0.27_{-0.38}^{+0.47}$ & $0.060_{-0.078}^{+0.031}$ & $0.06_{-0.11}^{+0.15}$ & 0.050\\ \vspace{0.5em}
$\alpha$ & $0.32_{-0.13}^{+0.24}$ & $0.32_{-0.31}^{+0.26}$ & $0.37_{-0.11}^{+0.25}$ & $0.37_{-0.34}^{+0.26}$ & 0.22\\ \vspace{0.5em}
$\beta$ & $0.80_{-0.76}^{+0.19}$ & $0.8_{-0.80}^{+1.3}$ & $1.19_{-1.19}^{+0.11}$ & $1.2_{-1.2}^{+2.4}$ & 1.09\\ \vspace{0.5em}
$w$ & $-1.034_{-0.011}^{+0.034}$ & $-1.034_{-0.042}^{+0.034}$ & $-1.029_{-0.010}^{+0.029}$ & $-1.029_{-0.043}^{+0.029}$ & -1.047\\ \vspace{0.5em}
$\Omega_{\rm b} h^2$ & $0.02227_{-0.00016}^{+0.00016}$ & $0.02227_{-0.00032}^{+0.00032}$ & $0.02226_{-0.00015}^{+0.00015}$ & $0.02226_{-0.00030}^{+0.00030}$ & 0.02227\\ \vspace{0.5em}
$\Omega_{\rm c} h^2$ & $0.139_{-0.022}^{+0.012}$ & $0.139_{-0.030}^{+0.034}$ & $0.1235_{-0.0056}^{+0.0027}$ & $0.124_{-0.0078}^{+0.010}$ & 0.1227\\ \vspace{0.5em}
$H_0 ~[{\rm km~ s^{-1} Mpc^{-1}}]$ & $66.6_{-1.7}^{+2.1}$ & $66.6_{-3.7}^{+3.6}$ & $67.94_{-0.74}^{+0.73}$ & $67.9_{-1.5}^{+1.5}$ & 68.74\\ \vspace{0.5em}
$\sigma_8$ & $0.757_{-0.047}^{+0.071}$ & $0.76_{-0.12}^{+0.11}$ & $0.810_{-0.013}^{+0.015}$ & $0.810_{-0.029}^{+0.028}$ & 0.812\\
\hline
\end{tabular}
\caption{\label{table:result2}Constraints on the parameters in the model $w>-1$.}
\begin{tabular}{cccccc}
\hline  \hline
& {\it Planck} & {\it Planck} & ~{\it Planck}+BAO+RSD+SNe~ & ~{\it Planck}+BAO+RSD+SNe~ & ~{\it Planck}+BAO+RSD+SNe~ \\
Parameter & $68\%$ C.L. & $95\%$ C.L. & $68\%$ C.L. & $95\%$ C.L. & Best fit \\ \hline\\ \vspace{0.5em}
$\lambda$ & $-0.1_{-3.4}^{+4.1}$ & $-0.1_{-9.5}^{+8.9}$ & $10_{-75}^{+76}$ & $10_{-190}^{+190}$ & 7\\ \vspace{0.5em}
$\alpha$ & $0.16_{-0.16}^{+0.047}$ & $0.16_{-0.16}^{+0.22}$ & $0.13_{-0.13}^{+0.040}$ & $0.13_{-0.13}^{+0.19}$ & 0.031\\ \vspace{0.5em}
$\beta$ & $24_{-11}^{+11}$ & $24_{-17}^{+16}$ & $31.8_{-4.7}^{+8.2}$ & $32_{-12}^{+8}$ & 34.6\\ \vspace{0.5em}
$w$ & $-0.9869_{-0.0083}^{+0.0028}$ & $-0.987_{-0.011}^{+0.016}$ & $-0.9901_{-0.0041}^{+0.0021}$ & $-0.9901_{-0.0056}^{+0.0066}$ & -0.9910\\ \vspace{0.5em}
$\Omega_{\rm b} h^2$ & $0.02225_{-0.00016}^{+0.00015}$ & $0.02225_{-0.00032}^{+0.00032}$ & $0.02230_{-0.00015}^{+0.00015}$ & $0.02230_{-0.00030}^{+0.00030}$ & 0.02213\\ \vspace{0.5em}
$\Omega_{\rm c} h^2$ & $0.1194_{-0.0017}^{+0.0018}$ & $0.1194_{-0.0039}^{+0.0039}$ & $0.1189_{-0.0013}^{+0.0013}$ & $0.1189_{-0.0025}^{+0.0026}$ & 0.1193\\ \vspace{0.5em}
$H_0 ~[{\rm km~ s^{-1} Mpc^{-1}}]$ & $67.09_{-0.67}^{+0.69}$ & $67.1_{-1.4}^{+1.4}$ & $67.44_{-0.55}^{+0.54}$ & $67.4_{-1.1}^{+1.1}$ & 67.12\\ \vspace{0.5em}
$\sigma_8$ & $0.814_{-0.011}^{+0.012}$ & $0.814_{-0.025}^{+0.027}$ & $0.8138_{-0.0092}^{+0.0094}$ & $0.814_{-0.019}^{+0.019}$ & 0.8060\\
\hline
\end{tabular}
\end{center}
\end{table*}

Fig.~\ref{fig:constraints} shows the confidence contours for the parameters in the models $w<-1$ and $w>-1$. Tables 
\ref{table:result1},\ref{table:result2} summarize the best-fit parameters and $1\, \sigma$, $2\, \sigma$ bounds from the 
CMB data and CMB+BAO+RSD+SNe dataset. We find the constraints for the two models $w<-1$ and $w>-1$ differ from each other significantly. 
Besides the parameters in our model, we also show the constraints 
on some cosmological parameters, which are similar with the results of $\Lambda$CDM model in Ref.~\cite{2016A&A...594A..13P}, 
except that a larger dark matter fraction $\Omega_{\rm c} h^2$ and a lower $\sigma_8$ are preferred by the CMB data.

For the model $w<-1$, $\lambda \in (-0.05, 0.21)$ at $95\%$ confidence level, indicating the deviation from $\Lambda$CDM 
model cannot be too large.  And $\alpha \in (0.03, 0.63)$, $\beta \in (0, 3.6)$, $w \in (-1.072, -1)$ at $95\%$ confidence level imply that a 
non-zero value of $\alpha$ is preferred by the dataset. If we limit the power indices to integers or half-integers for physical reasons, 
then the preferred values are $\alpha = 0.5$ and $\beta = 0, 0.5, 1$. The inclusion of large-scale structure (LSS) and SNe data 
improves the constraints of $\lambda$ significantly but has little impact on the other parameters.

For the model $w>-1$, $\beta$ and $\lambda$ are poorly constrained. 
The $2\, \sigma$ confidence regions for the other parameters are $\alpha \in (0, 0.32)$, $\beta \in (20, 43)$ and 
$w \in (-0.9957, -0.9835)$. Thus a non-zero and large value of $\beta$ is preferred by the data. In addition, $w$ obeys a nearly Gaussian distribution, peaking at $-0.99$. In this model, the viable parameter 
space is reduced by about $50 \%$ after including the LSS and SNe data.

\begin{figure*}[htb]
\includegraphics[width=0.5\textwidth]{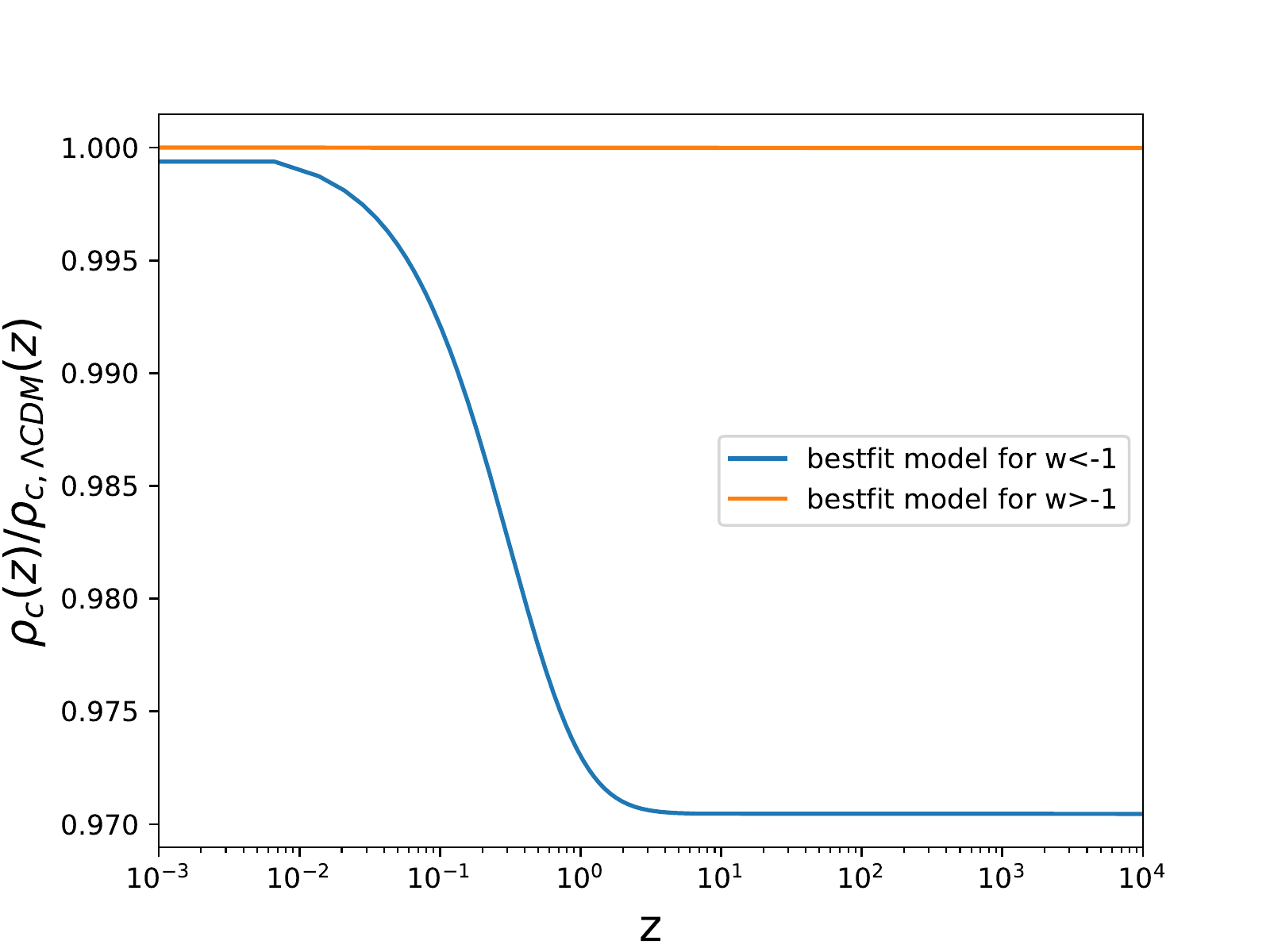}\includegraphics[width=0.5\textwidth]{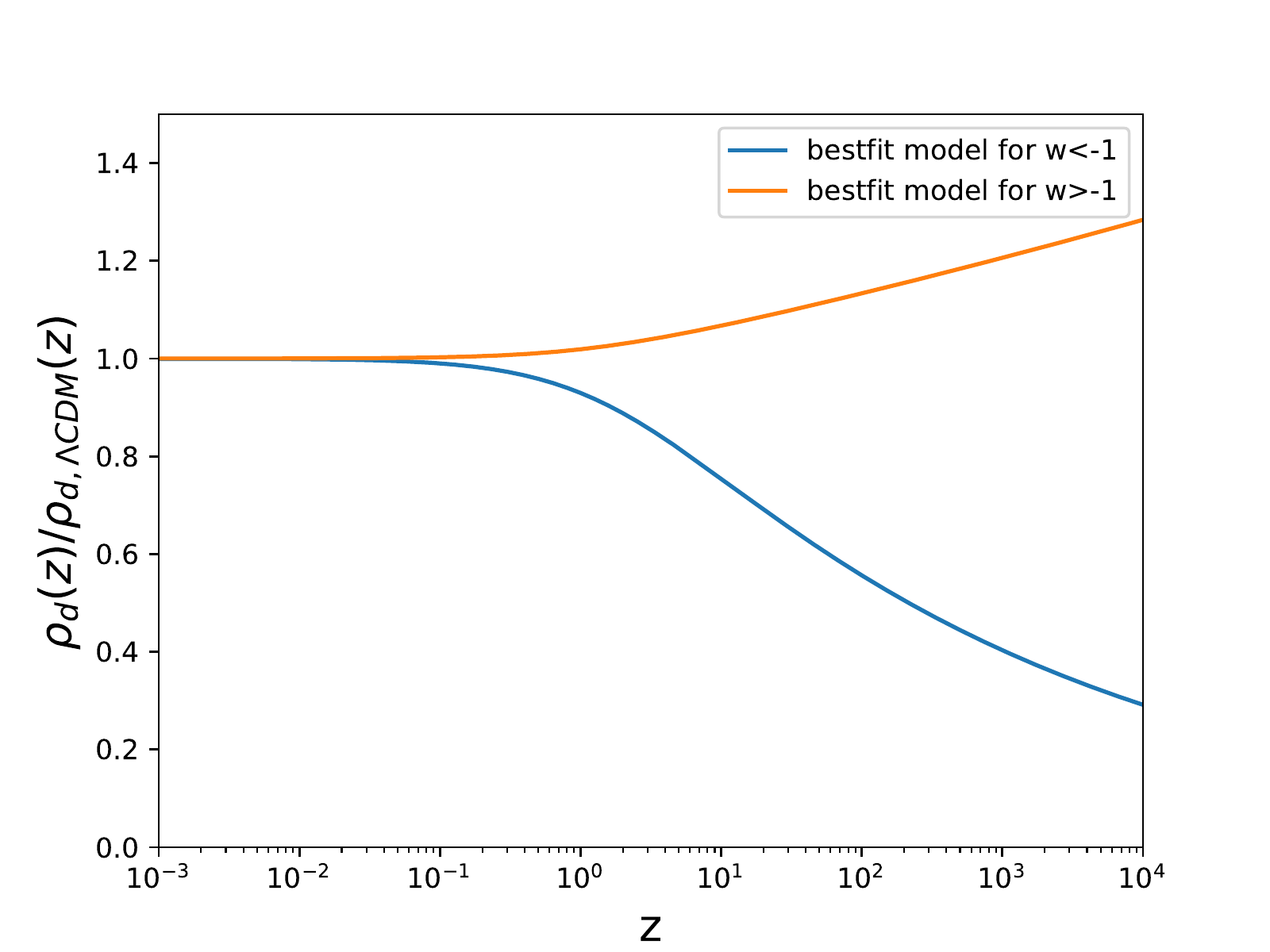}
\caption{The ratio of densities in the interacting models and $\Lambda \text{CDM}$ model for dark matter ({\it left} panel) 
and dark energy ({\it right} panel) with respect to the redshift $z$. We choose the best-fitting models preferred by the 
CMB+BAO+RSD+SNe data set for the models $w<-1$ and $w>-1$.}
\label{fig:back}
\end{figure*}

\begin{figure}[htb]
\includegraphics[width=0.5\textwidth]{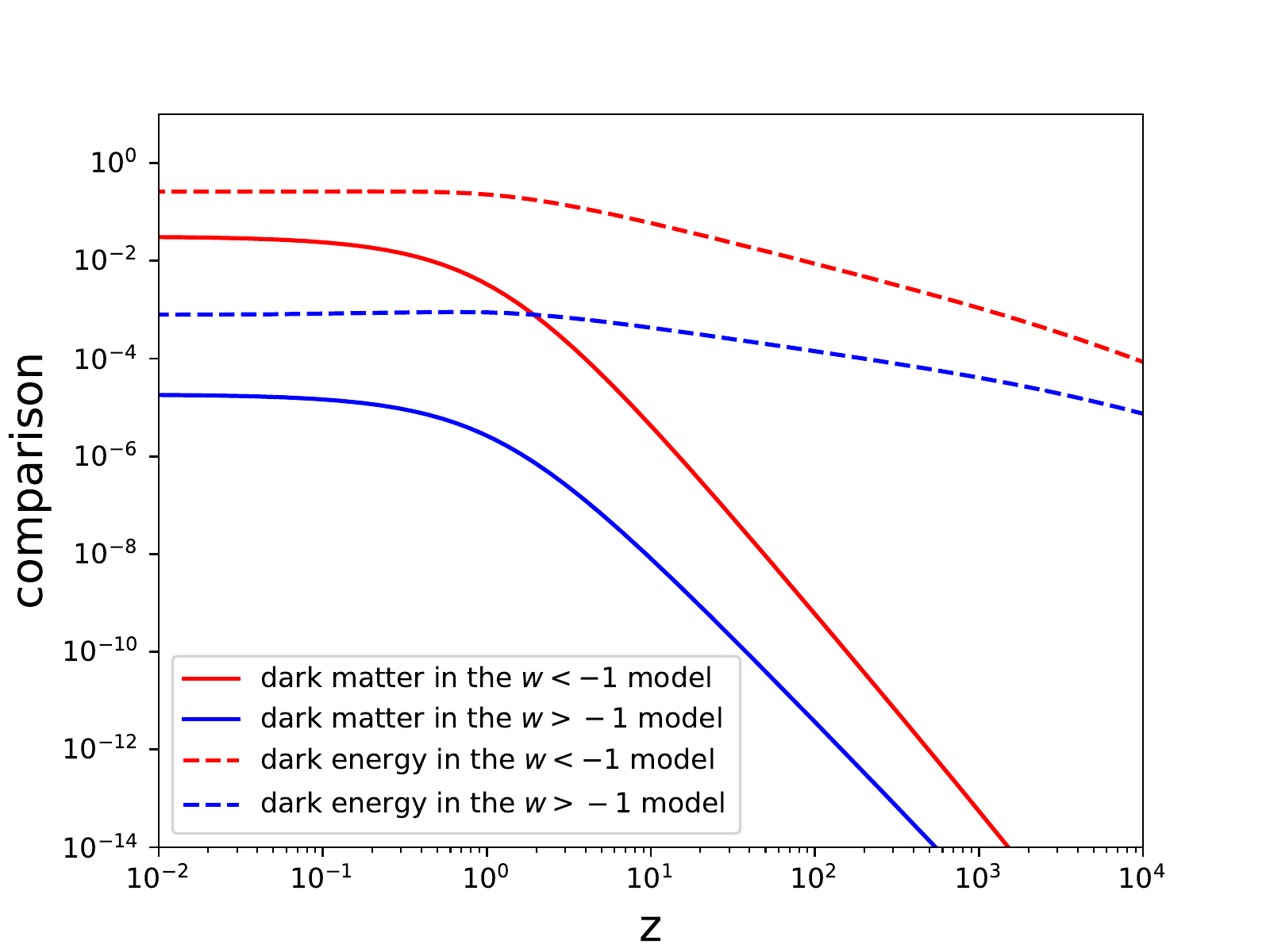}
\caption{The ratio of interaction term $a \gamma \rho_{\rm c}^{\alpha} \rho_{\rm d}^{\beta}$ and expansion term $3 a H \rho_{\rm c}$ 
for dark matter ($3 a (1 + w) H \rho_{\rm d}$ for dark energy) in the same best-fitting models as Fig.~\ref{fig:back}.}
\label{fig:comparison}
\end{figure}

In the following part, we take the best-fitting models as examples to discuss the background, perturbations and thermodynamics 
evolutions in the existence of interaction. Fig.~\ref{fig:back} shows the evolutions of densities for dark matter and dark energy, compared 
with the $\Lambda$CDM model. In both best-fitting models, $\lambda$ is positive, so the curves on the left panel have negative slopes,
suggesting dark energy is converted to dark matter. In the $w<-1$ model, the ratio of densities for dark matter 
is enhanced by several percentage from $z \sim 1$ to the present day and the density 
of dark energy at $z = 10^4$ is about $30 \%$ of the present day value. 

In the $w > -1$ model, a large $\beta$ is favored. In fact, the limit of $\beta$ is comparable 
to its prior range. Greater $\beta$ value would also be allowed if we have set a higher prior. To understand this, 
note that the interaction term is
$a \gamma \rho_{\rm c}^{\alpha} \rho_{\rm d}^{\beta} = a H_0 \rho_{\rm cr} \lambda (\rho_{\rm c}/\rho_{\rm cr})^{\alpha} (\rho_{\rm d}/\rho_{\rm cr})^{\beta}$,
for large $\beta$, since $\rho_d<\rho_{cr}$, this term becomes  
negligibly small compared to the expansion terms, so that despite of the appearance, the model actually behaves very similar 
to non-interaction $\Lambda$CDM model. The density evolution of dark matter is nearly identical with $\Lambda$CDM model, while the 
dark energy density is nearly constant as in the case of cosmological constant. Furthermore, in this case a large value of $\lambda$ is allowed, as the 
$(\rho_d/\rho_{cr})^{\beta}$ factor suppressed the whole term to very small value.

To quantify the contribution of interaction to the density evolution, we compare the interaction term with expansion term in Fig. 
\ref{fig:comparison}. In the matter dominant era, the ratio can be approximately written as $a \gamma \rho_{\rm c}^{\alpha} \rho_{\rm d}^{\beta} / 3 a H \rho_{\rm c} 
\propto (1 + z)^{3 \alpha - 4.5}$, so the contribution of interaction to the density evolutions is only important at low redshifts (roughly $z < 1$).

\begin{figure*}[htb]
\centerline{\includegraphics[width=0.5\textwidth]{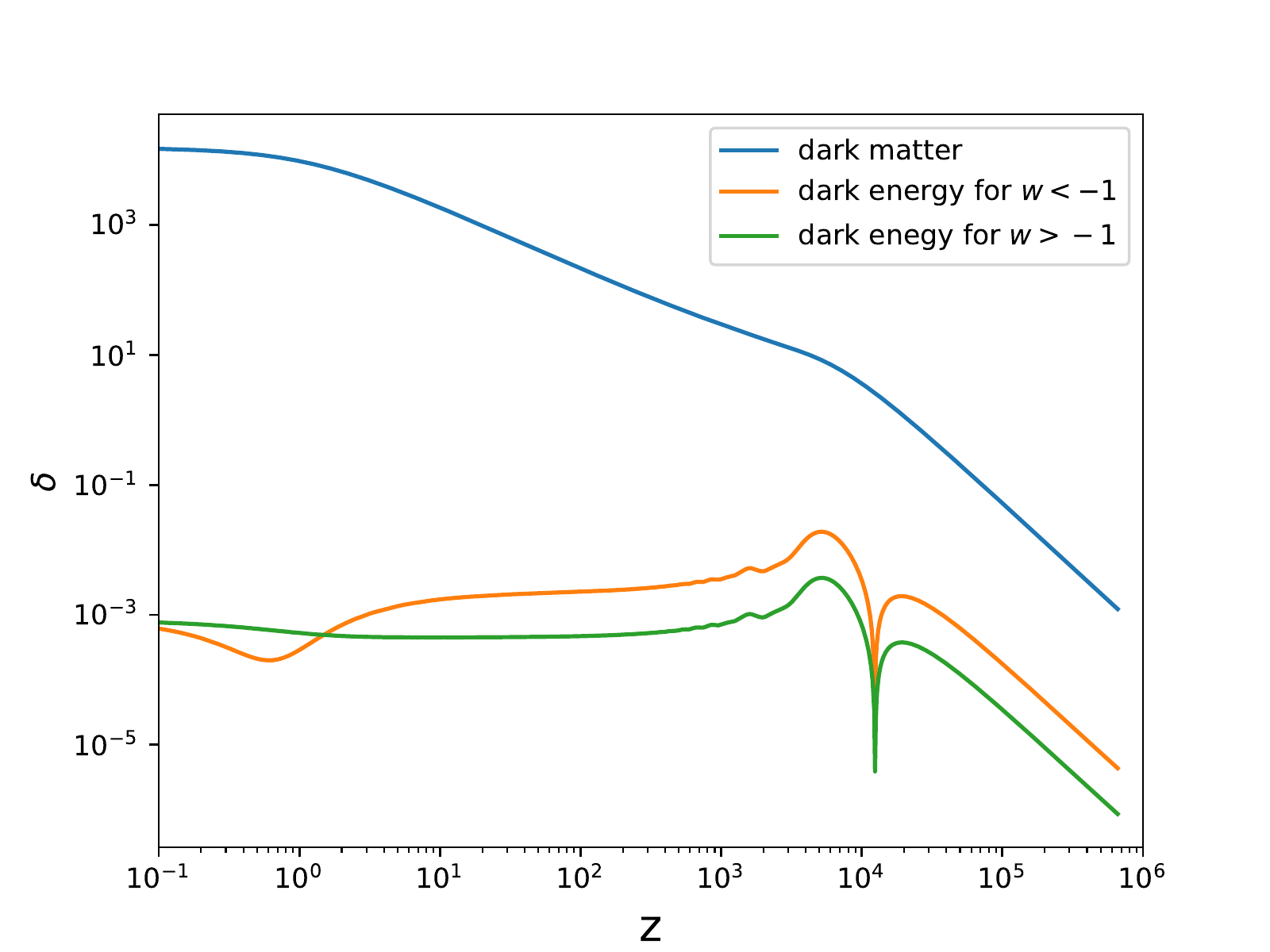}
\includegraphics[width=0.5\textwidth]{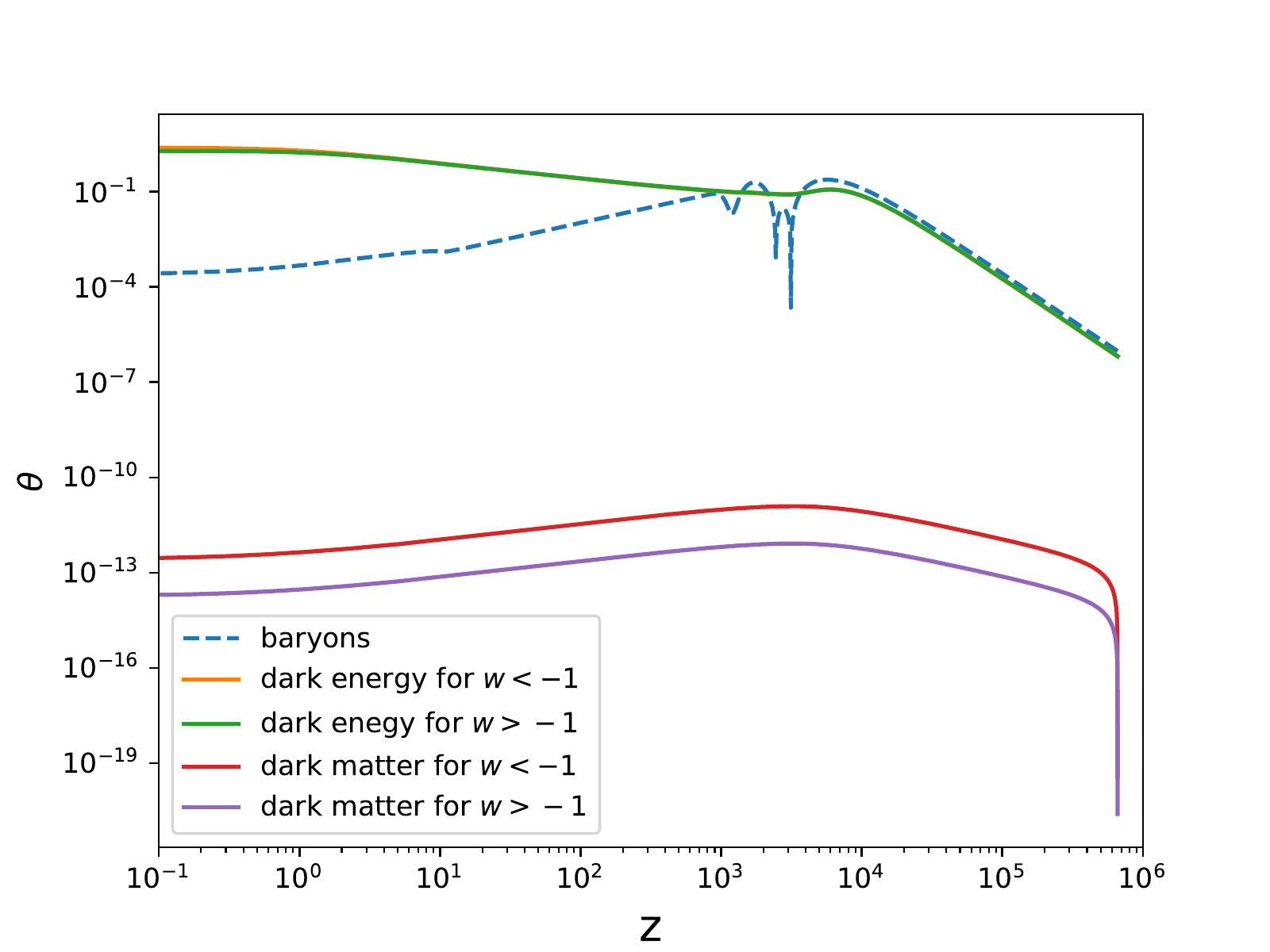}}
\caption{The density ({\it left} panel) and velocity ({\it right} panel) perturbations for baryons, dark matter and dark energy on the scale $k = 0.1~{\rm Mpc}^{-1}$ in the same 
models as Fig.~\ref{fig:back}. The evolutions of $\delta_{\rm c}$ are similar in different interacting models and $\Lambda$CDM model.}
\label{fig:perturbations}
\end{figure*}

\begin{figure}[htb]
\centerline{\includegraphics[width=0.5\textwidth]{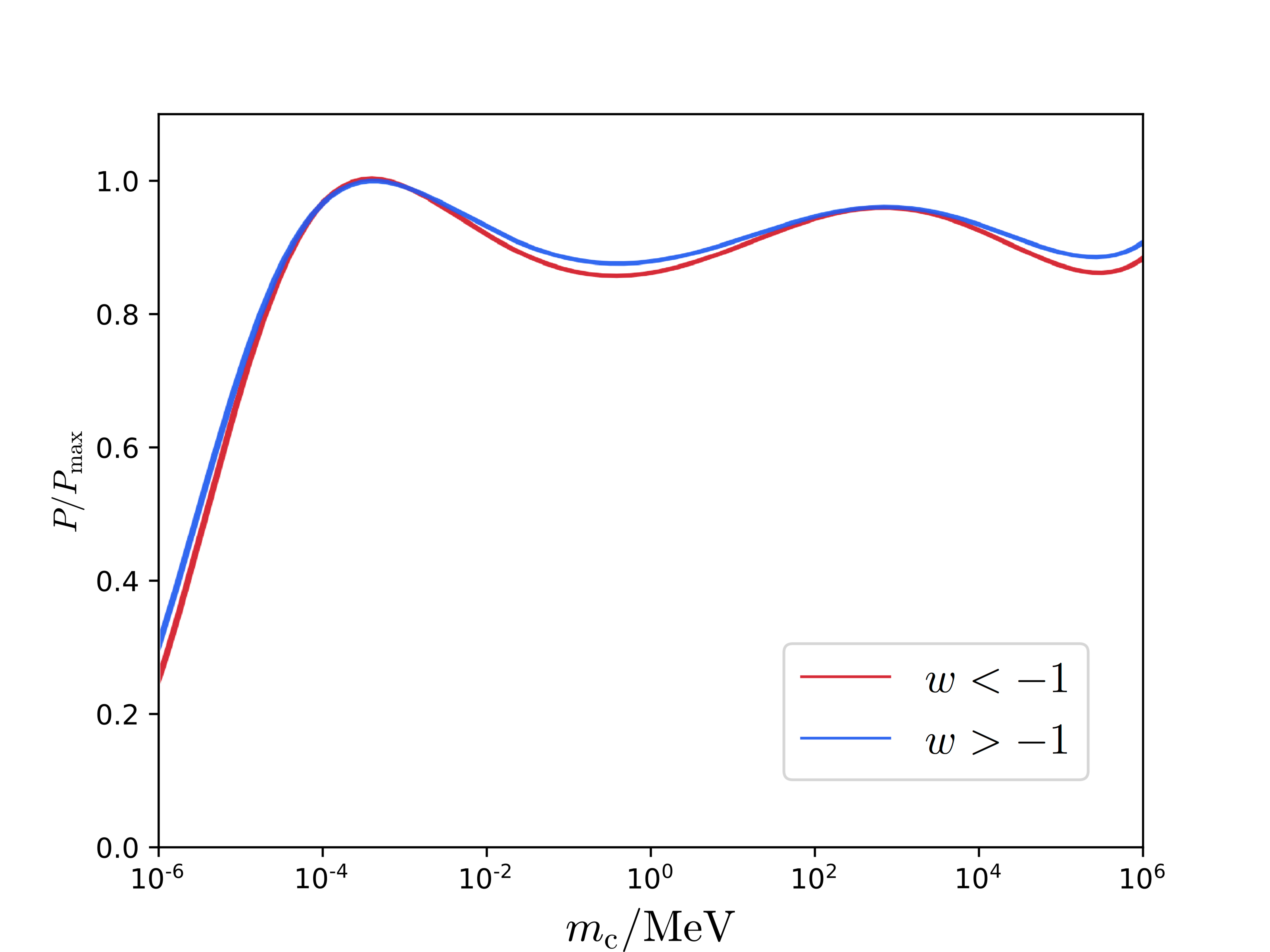}}
\caption{One-dimensional marginalised probability distribution for the parameter $m_{\rm c}$ constrained by CMB+BAO+RSD+SNe dataset.}
\label{fig:mass}
\end{figure}

\begin{figure*}[htb]
\includegraphics[width=0.5\textwidth]{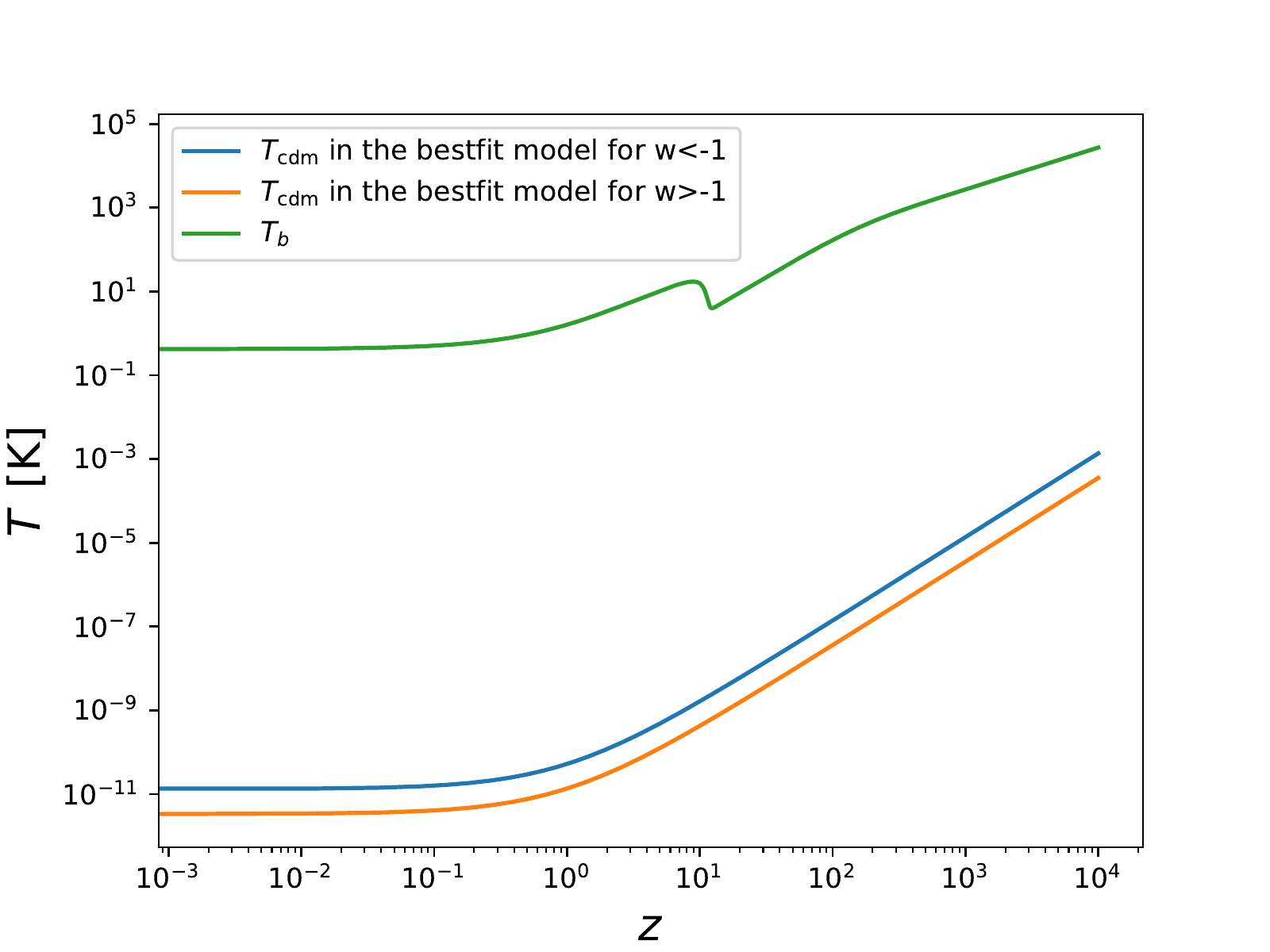}\includegraphics[width=0.5\textwidth]{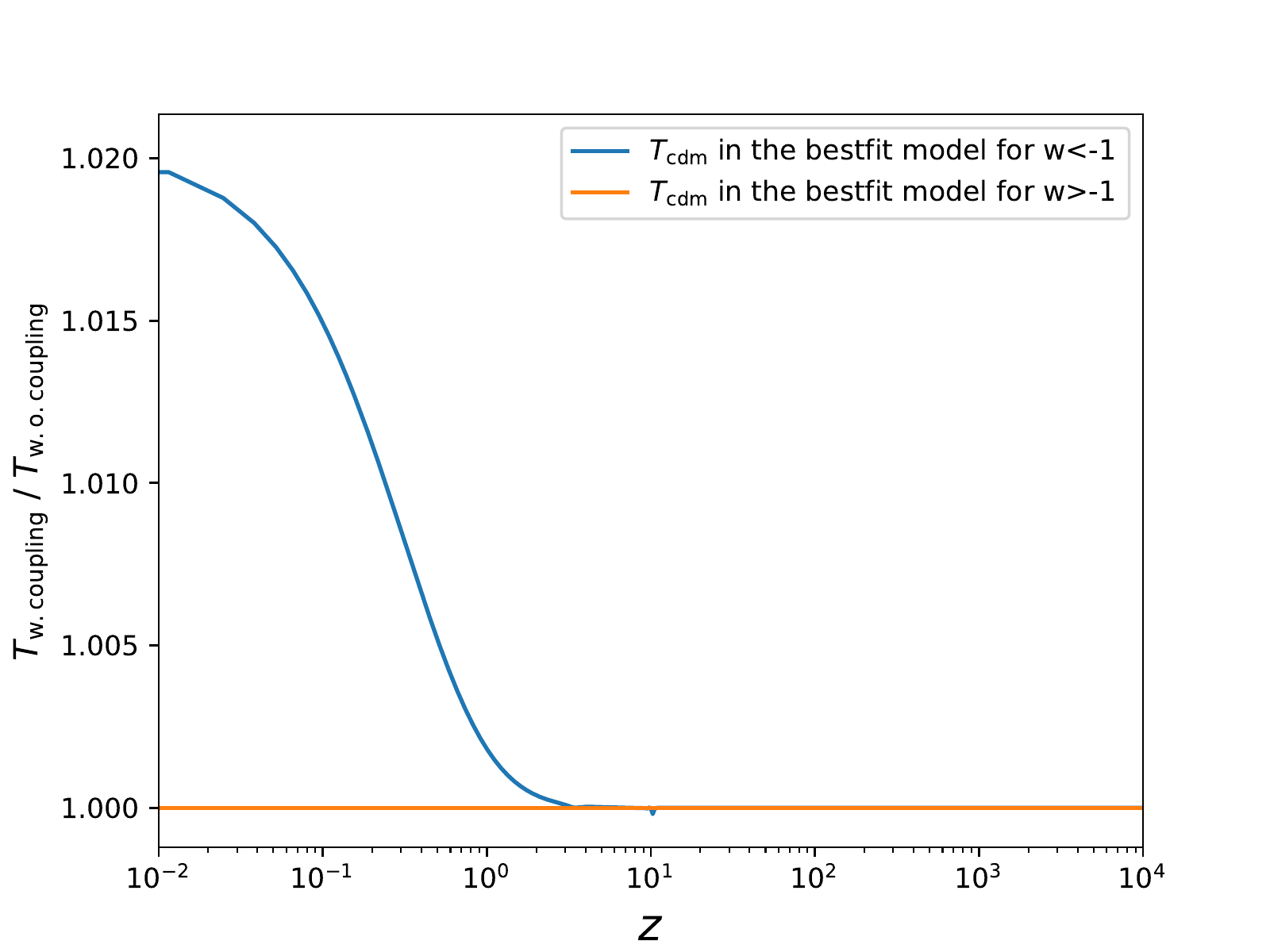}
\caption{{\it Left} panel: Temperature evolutions of baryons and dark matter in the best-fitting models as Fig.~\ref{fig:back}. {\it Right} panel: The ratio of temperature exactly evolved as Eq.~(\ref{eq:T}) 
and the temperature in the case without coupling (cooling adiabatically) under the same initial condition.}
\label{fig:temperature}
\end{figure*}

The cosmological perturbations are shown in Fig.~\ref{fig:perturbations}. We only plot the density perturbation of dark matter in the $\Lambda$CDM model, 
while the evolution in the interacting model could have several percent deviation. 
For the dark energy, the density perturbation which is several orders smaller than $\delta_{\rm c}$ grows and then keeps 
stable since $z \sim 10^4$. In the $\Lambda$CDM model, one usually sets $\theta_{\rm c} = 0$ in the synchronous gauge. In the interacting models, 
$\theta_{\rm c}$ has a tiny value. However, for dark energy, $\theta_{\rm d}$ is extremely large, even several orders larger than $\theta_{\rm b}$.
We note that to calculate $\dot{\theta}_{\rm c}$, the term $k^2 c_{\rm c}^2 \delta_{\rm c}$ is in the same order of $a H \theta_{\rm c}$ and even dominates in the 
range $0< z < 10^4$ and $0.01\,{\rm Mpc}^{-1} < k < 1\,{\rm Mpc}^{-1}$. So this term cannot be dropped. But $\theta_{\rm c}$ is really small and hence 
different choices of $m_{\rm c}$ have little impact on the evolution of the universe. We plot the posterior distribution of $m_{\rm c}$ in Fig.~\ref{fig:mass} and the 
distribution is nearly flat within the range $100\,{\rm eV}$--$1\,{\rm TeV}$. We have checked that for different choices of $m_{\rm c}$ in this range, the difference of 
the CMB power spectrum is smaller than $0.001\%$. The temperature evolutions of dark matter are shown in Fig.~\ref{fig:temperature}.
Dark matter cools nearly adiabatically until $z \sim 1$ and the temperature is enhanced due to the interaction by about $2 \%$ in the $w < -1$ model.
According to Eq.~(\ref{eq:T}), if $\gamma > 0$ dark matter is heated by dark energy and otherwise cooled.

\begin{figure}[htb]
\includegraphics[width=0.5\textwidth]{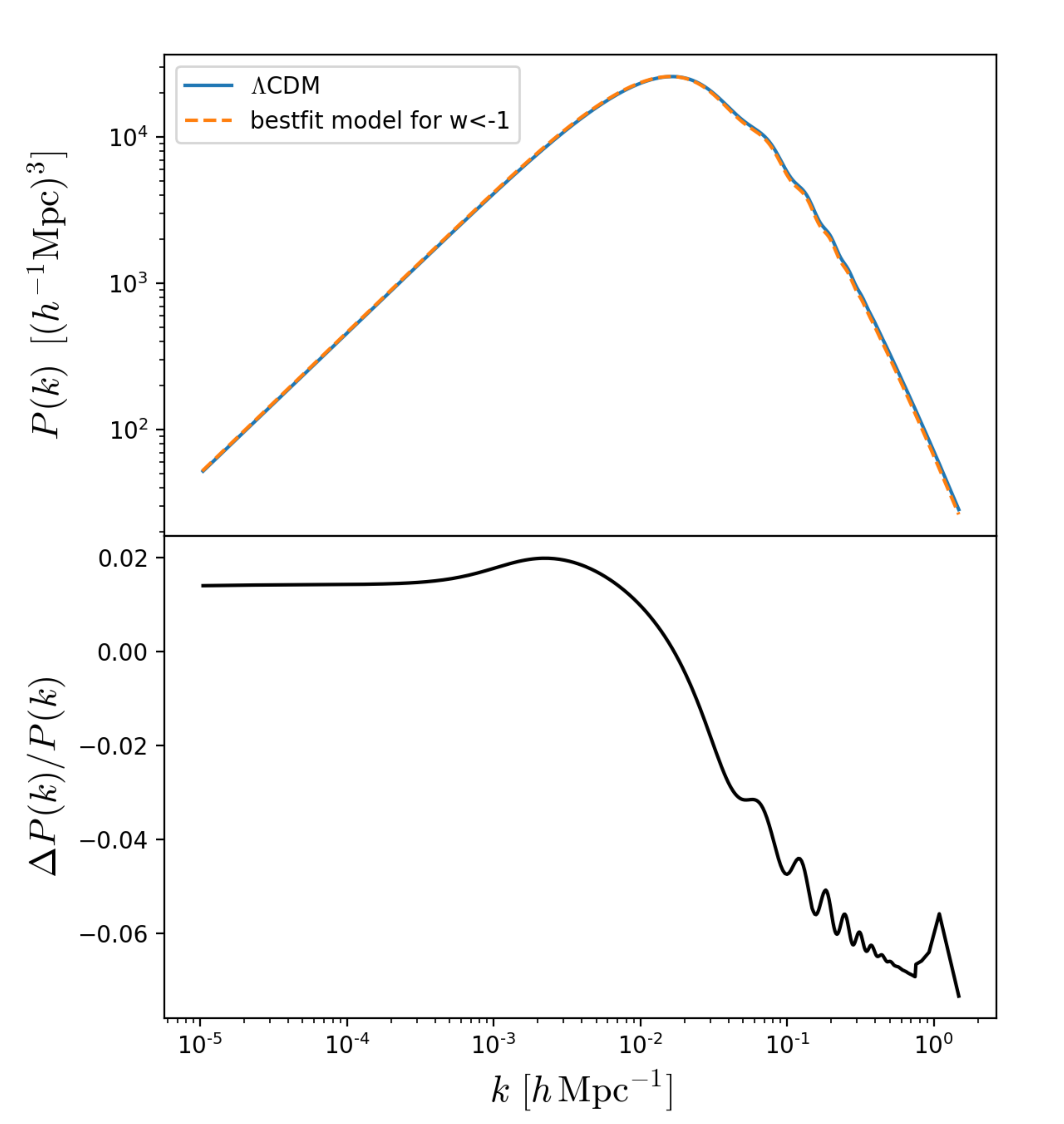}
\caption{{\it Top} panel: Linear matter power spectrum at $z = 0$ in the best-fitting models as Fig.~\ref{fig:back}. {\it Bottom} panel: The relative difference of 
matter power spectrum in the interacting model and $\Lambda$CDM model.}
\label{fig:pk}
\end{figure}

The linear matter power spectrum at present is shown in Fig.~\ref{fig:pk}. As illustrated in Fig.~\ref{fig:perturbations}, the dark matter perturbations
in the interacting model is similar with $\Lambda$CDM model. So the linear power spectrum makes little difference. The modification to the 
nonlinear power spectrum is much more significant and we will discuss it in the future paper.

\begin{figure}[htb]
\includegraphics[width=0.5\textwidth]{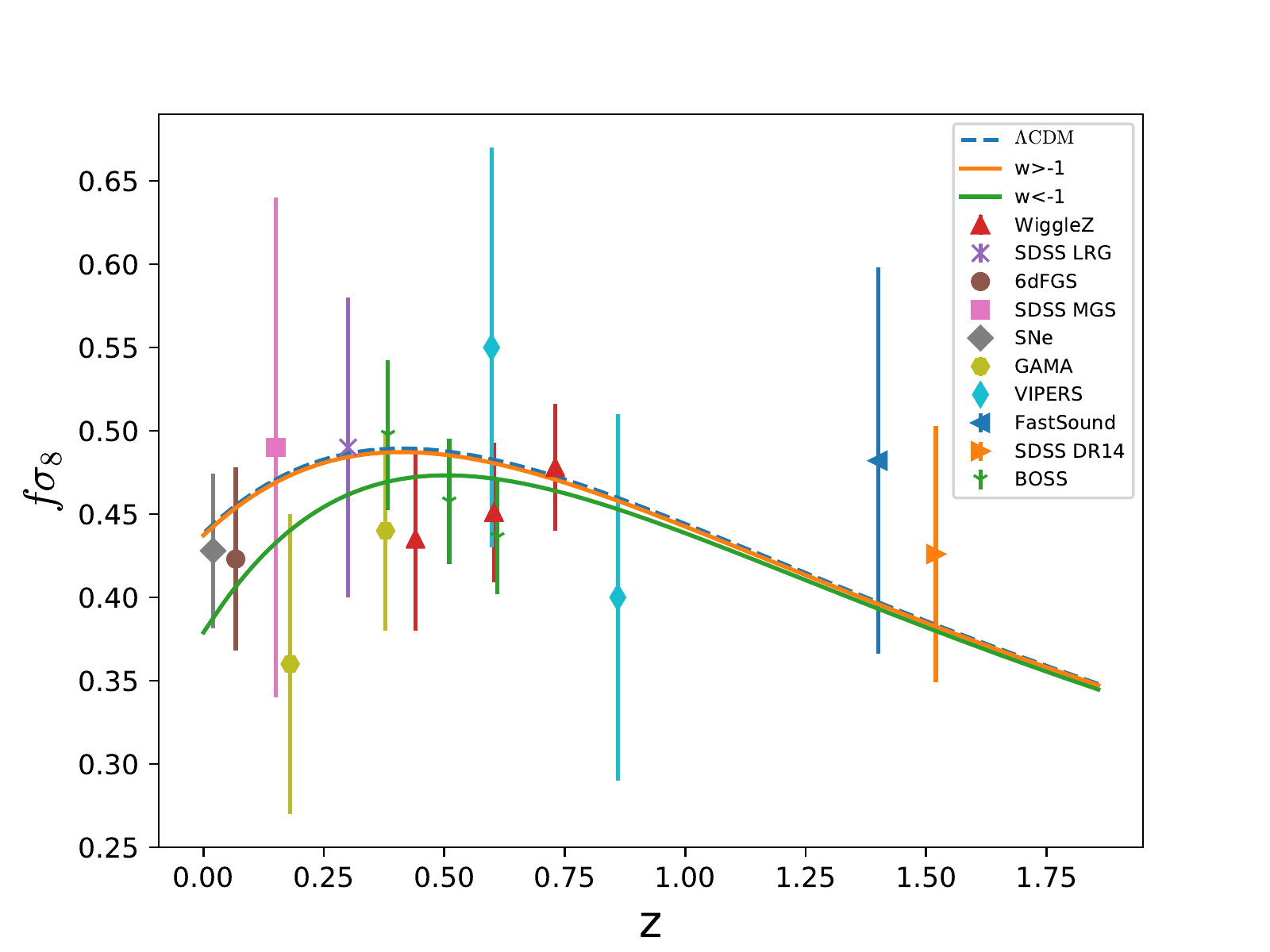}
\caption{$f(z) \sigma_8(z)$ evolutions with respect to the redshift in the best-fitting models as Fig.~\ref{fig:back} and $\Lambda$CDM model. 
The data points with error bar are measurements from the surveys listed in Table~\ref{table:rsd}. We shift some of the data point horizontally a bit ($\Delta z < 0.003$) to avoid overlap.}
\label{fig:rsd}
\end{figure}

We plot the evolutions of $f(z) \sigma_8(z)$ in our model and $\Lambda$CDM model in Fig.~\ref{fig:rsd}. $f \sigma_8 = a \der \sigma_8(z)/\der a$ 
reveals the time derivative of matter fluctuations on the scale $8~h^{-1}{\rm Mpc}$. Most measurements of $f(z) \sigma_8(z)$ at $z < 1$ are lower than 
that predicted by $\Lambda$CDM model. Coincidently, as discussed above, the interaction mainly influences the late time evolution 
of the universe (roughly $z < 1$). In the $w < -1$ model, the evolution of $f(z) \sigma_8(z)$ is similar with $\Lambda$CDM model at high redshift and decreases 
since $z < 1$. So the fit to data points in our model is much better than the $\Lambda$CDM model, alleviating the tension at low redshift. 
Quantitatively, we calculate the reduced chi-square $\chi_{\nu}^2 = \chi^2 / (n - m)$ to obtain the goodness of fit, where $n$ is the number of data points 
and m is the number of free parameters. We fix the cosmological parameters in different models and hence there are $5$ free parameters left. 
So $\chi_{\nu}^2$ is 0.56 for the $w < -1$ model and 0.75 for the $\Lambda$CDM model. Also, the curve predicted by the $w < -1$ model is 
within all the $1\, \sigma$ bounds of the measurements used in this plot.

\section{Summary}

In this paper, we consider the possible interaction between dark matter and dark energy and the impact on the evolution of universe. The interaction 
strength is proportional to some powers of the densities. We follow some previous work to derive the equations governing the evolutions of perturbations 
and thermodynamics in this model. And then we constrain the model using the latest cosmological data with the modified Boltzmann code.

For the $w > -1$ model,  the $2\, \sigma$ bounds are $\alpha \in (0, 0.32)$, $\beta \in (20, 43)$, 
$w \in (-0.9957, -0.9835)$. A very large value of $\beta$ is preferred by the data and $w$ obeys the Gaussian distribution, peaking at $-0.99$.

For the $w < -1$ model, the constraining results are $\lambda \in (-0.05, 0.21)$, $\alpha \in (0.03, 0.63)$, $\beta \in (0, 3.6)$ and $w \in (-1.072, -1)$ at $95\%$ confidence level. 
While the non-interacting case ($\lambda = 0$) could accommodate the current data, a positive $\alpha$ is preferred, indicating the coupling strength between dark sectors may 
depend on $\rho_{\rm c}^{\alpha}$. For the physical reasons, one can choose the indices as integers or half-integers and the allowed values 
within $1\, \sigma$ regions are $\alpha = 0.5$ and $\beta = 0, 0.5, 1$. The inclusion of LSS and SNe data improves the constraints of some parameters significantly.

The background density evolution of dark matter could deviate from $\Lambda$CDM model by several percent. We also consider the perturbations in 
the existence of interaction and the impact on the structure formation and redshift space distortion. 
The density perturbation of dark matter could have deviation from $\Lambda$CDM model by several percent. 
Notably, $\theta_{\rm d}$ is several orders larger than the velocity perturbation of baryons $\theta_{\rm b}$. We note that the term $k^2 c_{\rm c}^2 \delta_{\rm c}$ is dominant when 
calculating $\dot{\theta}_{\rm c}$ and hence could not be dropped. Because the velocity perturbation of dark matter is tiny, different choices of dark matter mass 
could hardly influence the results. In the $w < -1$ model, dark matter cools adiabatically, then heated due to the interaction by about $2 \%$ since $z \sim 1$.

The linear power spectrum in this model makes little difference and we will discuss the nonlinear power spectrum in the future paper. The observed 
$f(z) \sigma_8(z)$ values at $z < 1$ are mostly lower than that predicted by the $\Lambda$CDM model. As the effects of the interaction 
mainly appear at low redshift, the $w < -1$ model can alleviate the tension and fit the data points much better. Quantitatively, the reduced chi-square is 0.56 
for the $w < -1$ model and 0.75 for the $\Lambda$CDM model. In summary, this class of interacting model in the phantom region ($w < -1$) is physically 
plausible and could provide better fit to the current CMB data from {\it Planck}, BAO and RSD data from SDSS and Type-Ia supernovae from Pantheon samples.

\acknowledgments
G.C., F. W. and X.C. acknowledge the support of the NSFC through grant No. 11633004, 11473044, U1501501,
MoST through grant 2016YFE0100300,  the CAS through QYZDJ-SSW-SLH017 and XDB 23040100. Y.Z.M. acknowledges
 the support of NRF with grant no.105925, 109577, and 120378, and NSFC with grant no.~11828301.
J. Z. is supported by IBS under the project code, IBS-R018-D1.

\bibliography{ms}

\end{document}